\documentclass{revtex4-2}
\usepackage[utf8]{inputenc}
\usepackage{amsbsy}
\usepackage{amsopn}
\usepackage{amstext}
\usepackage{amsmath,amsthm,amsfonts,amssymb}
\usepackage[mathcal]{eucal}
\usepackage{mathrsfs}
\usepackage{braket}
\usepackage[english]{babel}
\usepackage{color}
\usepackage{esint}
\usepackage{graphicx}
\usepackage{float}
\usepackage{units}
\usepackage{blindtext}
\usepackage{hyperref}
\usepackage{textcomp}
\usepackage{orcidlink}

\DeclareGraphicsExtensions{.png,PNG,.pdf,.PDF}
\usepackage{hyperref}
\usepackage{slashed}
\newcommand{\ie}{\begin{equation}}
\newcommand{\fe}{\end{equation}}
\newcommand{\se}{\begin{eqnarray}}
\newcommand{\ff}{\end{eqnarray}}
\begin{document}

\title{Comment on ``Light deflection with torsion effects caused by a spinning cosmic
string''}
\author{R. R. S. Oliveira\,\orcidlink{0000-0002-6346-0720}}
\email{rubensrso@fisica.ufc.br}
\affiliation{Departamento de F\'isica, Universidade Federal do Cear\'a (UFC), Campus do Pici, C.P. 6030, Fortaleza, CE, 60455-760, Brazil}


\date{\today}

\begin{abstract}

In this comment, we showed that the line element (``metric'') worked by Jusufi (Eur. Phys. J. C 76: 332, 2016) is incorrect since he worked with a spinning cosmic string with torsion in $(2+1)$-dimensions (and in polar coordinates), which cannot happen. Indeed, as in $(2+1)$-dimensions there are no screw dislocations or screw torsion (a consequence of the absence of the third spatial dimension, i.e., of the $z$-axis), and being a cosmic string with torsion a type of screw dislocation, but of cosmic origin (i.e., are cosmic dislocations), implies that a $(2+1)$-dimensional cosmic string does not have/carry torsion (since $dz=0$). Therefore, the line element worked by Jusufi is physically incoherent/unacceptable and inappropriate for the research/study.

\end{abstract}

\maketitle

\section{Introduction}

In a paper published in the European Physical Journal C, Jusufi \cite{Jusufi} used a new geometrical method introduced by Werner to find the deflection angle in the weak limit approximation by a spinning cosmic string in the context of the Einstein–Cartan (EC) theory of gravity. To do such a study, the author used the line element (``metric'') of a spinning string cosmic with torsion in $(3+1)$-dimensions with cylindrical coordinates. Later, the author adopted the spherical coordinate system, where he fixed the polar angle as being $\theta=\pi/2$, that is, it defined/chose the equatorial plane to eliminate the cross-terms. In this way, the author started by adopting the String-Randers optical metric, then he applied the Gauss–Bonnet theorem to the optical geometry and derived the leading terms of the deflection angle in the equatorial plane. The author also showed that the light deflection is affected by the intrinsic spin of the cosmic string and torsion. In particular, this paper is well written and, in some ways, brings interesting results about how the parameters of the cosmic string (spin and torsion) affect the deflection angle of light.

However, analyzing in detail some papers in the literature on the theory of topological defects and, in special, the screw dislocation spacetime, which carries screw torsion or simply torsion (a cosmic string with torsion is based on this), we note that the line element (``metric'') worked by Jusufi \cite{Jusufi}, given by Eq. (5) of his paper, is incorrect and, therefore, is physically incoherent/unacceptable (or inappropriate for the research/study). In this way, the objective of the present comment is to show that working with the line element of a spinning cosmic string with torsion in $(2+1)$-dimensions (and in polar coordinates) in the form presented by Jusufi \cite{Jusufi} is a mistake. In particular, this error occurred because he fixed $\theta=\pi/2$ to obtain the equatorial plane, which is basically the polar plane. In other words, the line element of Jusufi \cite{Jusufi} became a $(2+1)$-dimensional line element in polar coordinates ($z=0$). In fact, according to Jusufu \cite{Jusufi}, to convert the cylindrical coordinate system to spherical coordinates, it is necessary to use some coordinate transformations, such as $z=r\cos{\theta}$ and $\rho=r\sin{\theta}$ ($r=\sqrt{x^2+y^2+z^2}$). However, for $\theta=\pi/2$, we have $z=0$ and $\rho=r=\sqrt{x^2+y^2}$, i.e., we have the polar plane (or the polar coordinates system).


\section{The screw dislocation spacetime and the Spinning Cosmic String spacetime}

Before we present the screw dislocation spacetime, it is important to highlight that a dislocation is a type of linear topological defect (of cosmic or condensed material origin) that carries torsion in its line element/metric (dislocations are well described using the concept of torsion), arises from a break of the translational symmetry, can be described by the Riemann–Cartan geometry (in cylindrical coordinates), and can be of two types: the spiral dislocation (where the torsion is introduced by modifying the $\rho$ and $\phi$ coordinates), and the screw dislocation (where the torsion is introduced by modifying the $z$ coordinate) \cite{da1,da2,da3,B,valanis,puntigam,katanaev}. So, in the case of the screw dislocations, they can manifest (or be classified) in two ways: the screw dislocation I (or (or type I screw dislocation), which is a distortion of a vertical line into a vertical spiral, and the screw dislocation II (or type II screw dislocation), which is a distortion of a circular curve into a vertical spiral, also known as the Katanaev–Volovich dislocation (we will soon see that a cosmic string with torsion is based on type II) \cite{da1,da2,da3,valanis,puntigam,katanaev}.

Now, with respect to the (type II) screw dislocation spacetime, such background is modeled by the following line element (``metric'') with signature $(-,+,+,+)$ and in relativistic cylindrical coordinates $(t,\rho,\varphi,z)$ \cite{B,bakke4,valanis,puntigam,katanaev,Moraes,Oliveira}
\begin{equation}\label{1}
ds^2=-dt^2+d\rho^2+\rho^2 d\varphi^2+(dz+\chi d\varphi)^2, \ \ (c=1),
\end{equation}
where $\chi$ is the torsion (constant real parameter) of the (torsional) topological defect, or, by using the crystallography language, is related to the Burgers vector $\Vec{b}=b\Vec{e}_z$, that is, $\chi=b/2\pi$. In this way, to incorporate/introduce torsion into the system, the $z$ coordinate (or $z$-axis) must be modified in the form $dz\to dz+\chi d\varphi$ \cite{Bezerra}. However, if the system is already two-dimensional ($dz=0$), or if the $z$ axis is deleted ($dz\to 0$), so now it does not make much sense to make such a modification to include torsion (that is, it only makes sense to have torsion in the system when $dz\neq 0$ in all circumstances) \cite{da1,da2,da3,B,bakke4,valanis,puntigam,katanaev,Moraes,Bezerra,Helliwell,Oliveira}. Besides, taking $dt=0$, we reduce the line element \eqref{1} to the line element of a usual screw dislocation in condensed matter or solid-state, that is, the line element \eqref{1} is the version or relativistic analog of a screw disclination in condensed matter (or solid-state) \cite{bakke4,valanis,Bezerra,Helliwell,puntigam,katanaev,Moraes,Bausch,Dzyaloshinskii}. 

On the other hand, with respect to the spinning cosmic string spacetime with torsion (or simply the spinning cosmic string spacetime), such a background is modeled by the following line element (a generalization of \eqref{1}) \cite{Bezerra,Helliwell,Ozdemir,Letelier,Gal}
\begin{equation}\label{2}
ds^2=-(dt+4GJ^t d\varphi)^2+d\rho^2+\alpha^2\rho^2 d\varphi^2+(dz+4GJ^z d\varphi)^2,  \ \ (c=1),
\end{equation}
where the constant parameters $J^t\geq 0$ and $J^z\geq 0$ are the intrinsic angular momentum (or spin) and the torsion (or ``Burgers vector'') of the cosmic string with a linear mass density $\mu\geq 0$, being $G$ the Newton’s gravitational constant, and $\alpha=1-4G\mu$ ($0<\alpha\leq 1$) is a disclination, topological or curvature parameter (due to the conical singularity along the $z$-axis, that is, $\alpha$ models a cosmic disclination), respectively. It is interesting to mention that for $J^t=J^z=0$, we obtain the line element of a usual, static or ordinary cosmic string \cite{oliveira1,oliveira2,oliveira3,oliveira4}; for $J^z=0$, we obtain the line element of a spinning cosmic string (thus, a more appropriate name of an object modeled by \eqref{2} would be a chiral cosmic string) \cite{Bezerra,Helliwell,oliveira5}; and for $J^t=0$, we obtain the line element generated by a cosmic (screw) dislocation (which would be the line element \eqref{1} with $\chi\equiv 4GJ^z$ and $\varphi\to\alpha\varphi$) \cite{Bezerra,Helliwell}. However, for $\mu=0$ (or $\alpha=1$), necessarily implies in $J^t=J^z=0$, that is, in the absence or complete non-existence of the cosmic string. Therefore, the chiral cosmic string is modeled by a conical disclination (described by conical curvature $\alpha$), by a time-like dislocation (described by spin $J^t$), and by a space-like dislocation (described by torsion $J^z$) \cite{Bezerra,Helliwell}. In particular, defining $a\equiv 4GJ^t$ and $\beta\equiv 4GJ^z$, we obtain the line element worked by Jusufi \cite{Jusufi}, given by Eq. (4) of his paper.

However, according to Ref. \cite{Juan}, where the authors worked with one type of dislocation (i.e., spiral/edge dislocation) applied to the electronic properties of the graphene and related two-dimensional systems, in $(2+1)$-dimensions, there are no screw dislocations simply because there is no third spatial dimension ($dz=0$), that is, as the antisymmetric part of the torsion (tensor) is rather related to time displacements, and as the time components of the torsion generated by a dislocation are zero, implies that this term vanishes always (i.e., in flat surfaces screw dislocations do not exist). For more (valuable) information about dislocations in graphene, we recommend Refs. \cite{Carpio,Lopez,Yazyev,Warner,Lehtinen,Butz,Robertson,Bonilla,Wang}. So, taking $dz=0$ in the system, must automatically imply in $J^z=0$ ($\beta=0$) or $\chi=0$ (absence of torsion); therefore, a cosmic string, modeled by \eqref{2}, or a screw dislocation, modeled by \eqref{1}, does not carry (have) torsion in $(2+1)$-dimensions (or in two spatial dimensions), but only in $(3+1)$-dimensions where $dz\neq 0$. Consequently, the line element worked by Jusufi \cite{Jusufi}, given by Eq. (5) of his paper, is incorrect and, therefore, is physically incoherent/unacceptable (or inappropriate for the research/study). On the other hand, Jusufi \cite{Jusufi} could have worked with a spinning cosmic string in $(2+1)$-dimensions, but modeled only by $a$ and $\alpha$ (with $\beta=0$), such as occurs in Ref. \cite{oliveira5} (but unfortunately, that was not the case).


\section{Final remarks}

In this comment, we showed that the line element (``metric'') worked by Jusufi \cite{Jusufi} is incorrect since he worked with a spinning cosmic string with torsion in $(2+1)$-dimensions, which cannot happen. Indeed, as in $(2+1)$-dimensions there are no screw dislocations or screw torsion (a consequence of the absence of the third spatial dimension), and being a spinning cosmic string with torsion a type of screw dislocation (but of cosmic origin), implies that a $(2+1)$-dimensional spinning cosmic string does not have/carry torsion (since $dz=0$). In this case, the cosmic sting is only modeled by two parameters: one that describes the conical curvature (given by $\alpha$) and the other that describes the spin (given by $J^t$). Therefore, the line element worked by Jusufi \cite{Jusufi} is physically incoherent/unacceptable and inappropriate for the research/study.

\section*{Acknowledgments}

\hspace{0.5cm}The author would like to thank the Conselho Nacional de Desenvolvimento Cient\'{\i}fico e Tecnol\'{o}gico (CNPq) for financial support.

\section*{Data availability statement}

\hspace{0.5cm} This manuscript has no associated data or the data will not be deposited. [Author’ comment: There is no data associated with this manuscript or no data has been used to prepare it.]

\end{document}